\documentclass[twocolumn,aps,amsmath,amssymb,showpacs,showkeys,floatfix,prl,superscriptaddress]{revtex4}

\usepackage{bbm,bm}
\usepackage[dvips]{graphicx}
\usepackage{color}
\usepackage{amsmath,amssymb,latexsym}

\renewcommand{\ol}[1]{\overline{#1}}
\newcommand{\micron}{\mu\mathrm{m}}

\newcommand{\F}{\mathbf{F}}
\newcommand{\K}{\mathbf{K}}
\renewcommand{\P}{\mathbf{P}}
\newcommand{\T}{\mathbf{T}}
\newcommand{\e}{\mathbf{e}}
\newcommand{\q}{\mathbf{q}}
\renewcommand{\r}{\mathbf{r}}
\renewcommand{\v}{\mathbf{v}}

\begin{document}

\title{Load response of shape-changing microswimmers\\ scales with their swimming efficiency} 

\author{Benjamin M. Friedrich}
\email{benjamin.m.friedrich@tu-dresden.de}
\affiliation{cfaed, TU Dresden, Dresden, Germany}

\date{\today}



\begin{abstract}
External forces acting on a microswimmer can feed back on its self-propulsion mechanism.
We discuss this load response for a generic microswimmer that swims by cyclic shape changes.
We show that the change in cycle frequency is proportional to the
Lighthill efficiency of self-propulsion.
As a specific example, we consider Najafi's three-sphere swimmer.
The force-velocity relation of a microswimmer
implies a correction for a formal superposition principle for active and passive motion. 
\end{abstract}

\maketitle
 
Microswimmers that periodically change their shape can swim actively in a fluid.
For example, biological cells such as sperm cells or motile green alga
are propelled in a fluid by long, slender cell appendages known as cilia and flagella, 
which perform regular bending waves \cite{Gray:1928}. 
At the relevant length and time scales of microswimming,
inertia is negligible and propulsion relies solely on viscous forces, 
corresponding to a regime of low Reynolds numbers \cite{Lauga:2009,Elgeti:2015}.

The fluid-structure interaction between a shape-changing microswimmer 
and the viscous fluid is bidirectional:
active shape-changes set the surrounding fluid in motion; 
conversely, hydrodynamic friction forces feed back on the active propulsion mechanism of the microswimmer
{\color{black}
and can change speed and shape of its swimming stroke.
} 
This load response becomes important when the swimmer is subject to an external force, 
and has implications for cargo transport and interactions between several microswimmers,
as well as for swimming in fluids of different viscosities. 
Furthermore, the load response may provide insight into the active propulsion mechanism itself.

Previous theoretical work considered shape-changing microswimmers towing a load \cite{Golestanian:2008,Golestanian2008b,Raz2007}.
In these studies, shape and timing of the swimming stroke was prescribed.
Other authors have formulated dynamic equations for the swimming stroke of microswimmers 
that employ active driving forces 
\cite{Gunther:2008,Friedrich:2012c,Polotzek:2013,Felderhof2014_3sphere,Pande:2015}. 
In this case, the speed of the swimming stroke depends on the external load.
{\color{black}
Experiments showed that the instantaneous phase speed of beating flagella 
indeed changes as a function of fluid viscosity \cite{Brokaw:1966,Woolley2001,Friedrich:2010}
or external flow velocity \cite{Wan2014rhythmicity,Quaranta:2015,Klindt:2016}.
} 

The feedback between hydrodynamic friction forces and the speed of the flagellar beat
is a prerequisite for the striking phenomenon of flagellar synchronization by hydrodynamic coupling \cite{Friedrich:2016}.
Collections of beating cilia and flagella can phase-lock their oscillatory bending waves 
\cite{Sanderson:1981,Ruffer:1998a,Riedel:2005a,Brumley:2014}.
Theory explains this phenomenon by hydrodynamic coupling between the cilia, 
where the hydrodynamic load acting on each `flagellar oscillator' 
depends on the phases of the other oscillators \cite{Gueron:1999,Vilfan:2006,Geyer:2013,Elgeti:2013}.

The load response of shape-changing microswimmers 
has also implications for a formal superposition principle for active and passive motion 
used in the literature \cite{Jekely:2008,Kummel2013,Kummel2014,tenHagen2015}.
This superposition principle states that active self-propulsion 
can be characterized by a fictitious propulsion force, 
such that the motion of a multi-component swimmer is characterized 
by the sum of the fictitious propulsion forces of its individual components \cite{Kummel2014}. 
This apparently simple superposition principle
provides formulas that formally resemble a force balance. 
The use of fictitious propulsion forces was critically commented on by Felderhof \cite{Felderhof2014}.
In fact, the use of fictitious propulsion forces seemingly contradicts the fact that 
self-propelled microswimmers do not exert any net force on the fluid as a consequence of Newton's third law.
{\color{black}
Furthermore, it is not clear if the formal superposition principle also holds 
in the presence of a load response of the active components.
}

Here, we highlight generic aspects of the load response of active swimmers,
and introduce a minimal model of a shape-changing microswimmer subject to external load.
This model abstracts from the intricate force-generation mechanisms of cilia and flagella, and other biological microswimmers.
We explicitly account for the conversion of energy during active motion,
from an energy reservoir into work performed on the surrounding fluid, 
and possibly dissipation inside the microswimmer itself. 
We make the idealizing assumption that the energy expenditure per shape-change cycle is independent of load.
This case corresponds to a maximal load response.
We briefly sketch a case of a load-dependent driving force.

Our contribution is two-fold:
First, we report a direct relationship between the 
response of a shape-changing microswimmer to external load, 
and its swimming efficiency.
The swimming efficiency 
generalizes the hydrodynamic propulsion efficiency of Lighthill \cite{Lighthill1952}, 
and measures the ratio between 
the power required to tow a passive swimmer of constant shape through the fluid, 
and the average rate of energy expenditure of the active, shape-changing swimmer \cite{Chen:2015}.
{\color{black}
Second, we propose that fictitious propulsion forces previously used to characterize active swimmers \cite{Kummel2014}
may be interpreted in terms of the constraining force required to constrain the active swimmer from moving.
}

The manuscript is structured as follows.
In the first section, we consider a generic shape-changing swimmer,
and its response to an external force parallel to its swimming direction.
We then discuss Najafi's three-sphere swimmer as a specific example \cite{Najafi:2004}, 
and review previous experiments of the load response of beating flagella \cite{Klindt:2016}.
In a last section, we generalize results to multi-component swimmers
and derive a superposition principles for passive and active motion 
valid in the case of a linear force-velocity relation of the active components. 

\paragraph{The force-velocity relation of a shape-changing swimmer.}
At the relevant length and time scales of microswimming, fluid flow is governed by the Stokes equation, 
$0=\eta \nabla^2 \mathbf{u}-\nabla p$, 
where $p$ and $\mathbf{u}$ denote the pressure and velocity field of the fluid, respectively, and $\eta$ its dynamic viscosity. 
We consider a microswimmer that swims in a viscous fluid by actively changing its shape in a cyclic fashion.
We parametrize the cyclic shape sequence of the swimmer by a $2\pi$-periodic variable $\varphi$.
For simplicity, we assume that motion of this swimmer is constrained along the $x$ axis.

The swimmer is thus described by two degrees of freedom only, phase $\varphi$ and position $x$.
Its motion follows from a balance of generalized forces 
in the sense of Lagrangian mechanics of dissipative systems \cite{Goldstein:mechanics,Polotzek:2013}
\begin{align}
\label{eq_motion}
F_\mathrm{ext} &= \Gamma_{xx} \dot{x} + \Gamma_{x\varphi}\dot{\varphi}, \\
\label{eq_motion2}
Q &= \Gamma_{\varphi x} \dot{x} + \Gamma_{\varphi\varphi}\dot{\varphi} + \kappa \dot{\varphi}. 
\end{align}
Here, $Q$ denotes a generalized active driving force, conjugate to $\varphi$, with unit of a torque.
The external force $\F_\text{ext}=F_\text{ext}\mathbf{e}_x$ shall act along the $x$ axis.
The generalized hydrodynamic friction coefficients $\Gamma_{ij}$ depend on the phase $\varphi$.
We have an Onsager-type relation $\Gamma_{x\varphi}=\Gamma_{\varphi x}$ \cite{Happel:hydro}.
In Eq.~(\ref{eq_motion2}), we have accounted for the possibility of internal friction 
of the active propulsion mechanism by a friction term $\kappa\dot{\varphi}$ 
with internal friction coefficient $\kappa\ge 0$.

While Eqs.~(\ref{eq_motion}) and (\ref{eq_motion2}) are general, 
it is instructive to consider as a specific example a microswimmer built from $n$ spheres.
A general motion of the spheres (comprising translations and rotations) 
is characterized by a grand hydrodynamic friction matrix $\mathbf{\Gamma}_0$
of dimensions $6n\times 6n$ \cite{Happel:hydro}.
We are interested in a cyclic shape change, 
corresponding to a cyclic sequence of relative geometric configurations of the spheres, 
parametrized by a phase variable $\varphi$.
The reduced hydrodynamic friction matrix $\mathbf{\Gamma}$ of dimensions $2\times 2$
with components $\Gamma_{xx}$, $\Gamma_{x\varphi}$, $\Gamma_{\varphi x}$, $\Gamma_{\varphi \varphi}$
can then be expressed as a contraction of the grand hydrodynamic friction matrix $\mathbf{\Gamma}_0$ as
$\mathbf{\Gamma}=\mathbf{L}\cdot\mathbf{\Gamma}_0\cdot\mathbf{L}^T$.
The matrix $\mathbf{L}$ specifies the velocities of the $n$ spheres
if either $\varphi$ changes, or the swimmer moves as a whole, 
see \cite{Polotzek:2013} for details. 
We now continue with the general treatment.

During its motion, the microswimmer dissipates energy at a rate
$\mathcal{R}_\mathrm{tot}=\mathcal{R}^{(h)}+\mathcal{R}^{(i)}$,
where 
\begin{equation}
\mathcal{R}^{(h)}=
F_\mathrm{ext}\dot{x} +
(\Gamma_{\varphi x}\dot{x}+ \Gamma_{\varphi\varphi}\dot{\varphi})\dot{\varphi}
\end{equation}
is the rate at which work is performed on the fluid, 
and
$\mathcal{R}^{(i)}=\kappa\dot{\varphi}^2$
is the rate of energy dissipation inside the swimmer itself.
The microswimmer depletes an internal energy store at rate 
$\mathcal{R}_\mathrm{int}=Q\dot{\varphi}$, 
which characterizes its active propulsion mechanism.
Energy conservation reads
\begin{equation}
\label{eq:R}
\mathcal{R}^{(h)}+\mathcal{R}^{(i)} =
\mathcal{R}_\mathrm{int} + \mathcal{R}_\mathrm{ext},
\end{equation}
where
$\mathcal{R}_\mathrm{ext}=F_\mathrm{ext}\dot{x}$ 
equals the rate at which the external force performs work on the system.

We introduce the time-averaged mobility coefficient $\mu=\langle \Gamma_{xx}^{-1} \rangle$
and the net swimming velocity $v_a=\Delta x/T$ of the swimmer in the absence of an external force.
Here, $\Delta x=x(T)-x(0)$ is the net displacement after one swimming cycle with duration $T$.
We define the swimming efficiency $\varepsilon_\mathrm{swim}$ of the microswimmer
as the ratio between 
(i) the hydrodynamic dissipation rate 
$\mathcal{R}_\mathrm{drag}=v_a F_a$ 
for dragging a passive particle with time-independent mobility $\mu$ through the fluid 
by an external force $F_a=v_a/\mu$ at constant speed $v_a$, 
and 
(ii) the average rate $\langle\mathcal{R}_\mathrm{int}\rangle$ of energy expenditure of the active swimmer
\begin{equation}
\label{eq:L}
\varepsilon_\mathrm{swim} = \frac{\mathcal{R}_\mathrm{drag}}{\langle\mathcal{R}_\mathrm{int}\rangle}.
\end{equation}
Following \cite{Chen:2015}, 
we can factor $\varepsilon_\mathrm{swim}$ as a product 
\begin{equation}
\varepsilon_\mathrm{swim}
=\varepsilon_\mathrm{hydro}\,\varepsilon_\mathrm{chem}, 
\end{equation}
where
$\varepsilon_\mathrm{hydro}=\mathcal{R}_\mathrm{drag} / \langle\mathcal{R}^{(h)}\rangle$
is the hydrodynamic propulsion efficiency of Lighthill \cite{Lighthill1952}, 
and 
$\varepsilon_\mathrm{chem}=\langle\mathcal{R}^{(h)}\rangle / \langle\mathcal{R}_\mathrm{int}\rangle$
represents a chemo-mechanical efficiency that 
characterizes the conversion of internal energy into work performed on the fluid.
Note that unlike \cite{Chen:2015}, 
our definition of the mechanical power output used in $\varepsilon_\mathrm{chem}$ 
considers only the work performed by the swimmer on the fluid, 
and not work performed by the swimmer on itself.
In the limit of large internal friction, $\kappa\gg\Gamma_{\varphi\varphi}$, 
we have $\varepsilon_\mathrm{chem}\sim 1/\kappa$. 
The internal friction coefficient $\kappa$ 
allows to interpolate between the case of 
a swimming stroke that is maximally susceptible to external forces ($\kappa=0$), 
and the case of a prescribed swimming stroke ($\kappa\gg\Gamma_{\varphi\varphi}$).

We assume in the following that the rate $\mathcal{R}_\mathrm{int}$ of energy expenditure 
does not depend on the phase $\varphi$ for the reference case $F_\mathrm{ext}=0$.
This assumption will simplify calculations and corresponds to an optimal swimming stroke.
For a given sequence of shapes, but variable driving protocol $Q(\varphi)$, 
the case of constant energy expenditure 
minimizes the total hydrodynamic dissipation during a swimming cycle
(if the cycle period $T_0=2\pi/\omega_0$ is held fixed in the optimization)
\cite{Alouges:2009}.
In this case, we can always find a phase parametrization of the swimming stroke 
such that the active driving force $Q$ does not depend on phase.
For $F_\mathrm{ext}=0$, we then additionally have that the phase speed is constant, $\dot{\varphi}=\omega_0$. 
The instantaneous swimming velocity in the absence of load is given by 
$v_0(\varphi)=-\Gamma_{x\varphi}\Gamma_{xx}^{-1}\omega_0$.

We now consider the case of a non-zero external force $F_\mathrm{ext}$.
In this case, the net swimming speed $v(F_\text{ext})$ can be written as a sum of three terms:
(i) the active propulsion velocity $v_a$ in the absence of an external force,
(ii) the passive drag velocity ${\mu} F_\text{ext}$ induced by the external force, and 
(iii) a load response $\Phi(F_\text{ext})$, 
which characterizes the feedback of the external force on the propulsion mechanism
\begin{equation}
\label{eq:phi}
v(F_\text{ext})=v_a+\mu F_\text{ext}+\Phi(F_\text{ext}).
\end{equation}
We now compute $\Phi(F_\text{ext})$.
From the equation of motion, Eq.~(\ref{eq_motion}), 
we obtain a force-velocity relation for the instantaneous phase velocity
\begin{equation}
\label{eq_force_velocity}
\dot{\varphi} = \omega_0 + \frac{v_0(\varphi)}{Q}F_\mathrm{ext}.
\end{equation}
Thus, for $F_\mathrm{ext}>0$, the swimming stroke speeds up during the effective stroke with forward motion $v_0(\varphi)>0$,
and slows down during the recovery stroke with $v_0(\varphi)<0$.

By averaging Eq.~(\ref{eq_force_velocity}) over one cycle, 
we find for the angular frequency $\omega$ of the swimming stroke under load
\begin{equation}
\label{eq:omega}
\omega \approx \omega_0 \left( 1 + \frac{W}{E} \right).
\end{equation}
Here, $W = \int_0^T \! dt\, F_\mathrm{ext} \dot{x} = F_\mathrm{ext}\Delta x$ 
denotes the work performed by the swimmer in the external force field during one swimming cycle,
while $E = \int_0^T \! dt\, \mathcal{R}_\mathrm{int} = 2\pi Q$ 
is the energy expended by the microswimmer.
Eq.~(\ref{eq:omega}) is valid to leading order in $F_\mathrm{ext}$, 
corresponding to the limit $\mu|F_\text{ext}|\ll v_0(\varphi)$,
for which a time average and a phase average are approximately equal.

The net displacement of a shape-changing microswimmer depends only on the sequence of its shape changes, 
but not on the timing of its swimming stroke, i.e.\ $\dot{\varphi}(\varphi)$.
Thus, by the superposition principle for Stokes flow, 
the net displacement of a swimmer under load equals 
its net displacement $\Delta x$ without load, 
plus the displacement $\mu F_\mathrm{ext}T$ that a passive particle with mobility $\mu$
subject to an external force $F_\mathrm{ext}$ would experience
during one cycle of duration $T=2\pi/\omega$.
Dividing by $T$, 
we obtain a linear force-velocity relationship for the net swimming velocity, valid for small external forces 
\begin{equation}
\label{eq:v}
v(F_\mathrm{ext})\approx v_a \left(1 + \frac{W}{E} \right) + \mu F_\mathrm{ext}.
\end{equation}
Note that the time-averaged mobility $\mu$ changes 
if the timing of the swimming stroke is altered by the external force, 
yet this effect introduces only higher-order terms.
The term $\mu F_\mathrm{ext}$ equals the 
velocity of a passive swimmer subject to an external force, 
and has been previously discussed by Golestanian \cite{Golestanian2008b}. 
The term $\Phi=v_a W/E$ is new and characterizes a feedback of the external force 
on the active propulsion mechanism itself. 

We can rewrite the load response $\Phi$ in terms of the swimming efficiency $\varepsilon_\mathrm{swim}$ defined in Eq.~(\ref{eq:L}). 
We thus obtain our main result that $\Phi$ is proportional to $\varepsilon_\mathrm{swim}$
\begin{equation}
\label{eq:phi2}
\Phi(F_\text{ext}) = \varepsilon_\mathrm{swim}\,\mu F_\text{ext}.
\end{equation} 
For larger external forces $F_\text{ext}$, higher-order terms become important. 
If we extrapolate our linear theory to large forces, 
the constraining force at which net motion of the microswimmer vanishes, 
reads $F_c\approx -F_a/(1+\varepsilon_\mathrm{swim})$, 
with $F_a=v_a/\mu$.
For even larger forces, 
the phase speed $\dot{\varphi}$ becomes zero
at critical stall forces $F_\mathrm{stall}^\pm = -\omega_0 Q/v^\pm$
according to Eq.~(\ref{eq:phi}).
This corresponds to stalling of the swimming stroke at a specific phase $\varphi=\varphi^\pm$ of the cycle.
Here, 
$v^+=v_0(\phi^+)=\min_\varphi v_0(\varphi)$ 
denotes the minimum of the instantaneous swimming velocity $v_0(\varphi)$,
and $v^-=v_0(\phi^-)$ its maximum. 

In our minimal model, we made the idealizing assumption that 
the active driving force is independent of load, 
which implies a constant energy expenditure $E=E_0$ per shape-change cycle.
In the remainder of this section, 
we briefly sketch an illustrative example of a load-dependent active driving force, 
given by 
\begin{equation}
\label{eq:Qlambda}
Q=Q_0 - \lambda \frac{v_0(\varphi)}{\omega_0} F_\mathrm{ext},\quad
0\le\lambda\le 1.
\end{equation}
{\color{black}
This particular choice of $Q=Q(F_\mathrm{ext})$ is motivated by the form of the force-velocity relation for the instantaneous phase, 
Eq.~(\ref{eq_force_velocity}),  and will give particularly simple results.
}
The case $\lambda=0$ corresponds to the case $Q=Q_0$ considered above.
Analogously to Eq.~(\ref{eq:omega}), 
we find for general $\lambda$ to leading order in $F_\mathrm{ext}$, 
$\omega \approx \omega_0[1+(1-\lambda) W/E_0]$ and
$\langle\mathcal{R}_\mathrm{int}\rangle \approx Q_0\omega_0 + (1-2\lambda) \mathcal{R}_\mathrm{drag}$,
where $E_0=2\pi Q_0$ denotes the energy expenditure per swimming cycle in the absence of load.
The response to an external force $F_\mathrm{ext}$ 
that acts in the opposite direction as the swimmer moves
(i.e.\ $W=F_\mathrm{ext}\Delta x<0$, corresponding to ``head wind'')
is summarized in Table \ref{tab:lambda}.
While the load dependence of the cycle frequency is maximal for $\lambda=0$, it vanishes for $\lambda=1$.
Recent experiments indicate that the rate of ATP hydrolysis and thus the energy dissipation rate of beating flagella 
is rather insensitive to mechanical load \cite{Chen:2015}, 
which would correspond to an intermediate case $\lambda\approx 1/2$.

\begin{table}[h]
\begin{center}
\begin{tabular}{c|c|c|c|c}
load          & active  & cycle     & rate of     & energy      \\
dependence    & driving & frequency & energy      & expenditure \\
driving force & force   &           & expenditure & per cycle   \\
\hline
$\lambda=0$   & $Q=Q_0$                & $\omega<\omega_0$ & $\langle\mathcal{R}_\mathrm{int}\rangle <Q_0\omega_0$  & $E=E_0$ \\
$\lambda=1/2$ & $\langle Q\rangle>Q_0$ & $\omega<\omega_0$ & $\langle\mathcal{R}_\mathrm{int}\rangle \approx Q_0\omega_0$  & $E>E_0$ \\
$\lambda=1$   & $\langle Q\rangle>Q_0$ & $\omega=\omega_0$ & $\langle\mathcal{R}_\mathrm{int}\rangle >Q_0\omega_0$        & $E>E_0$ \\
\hline
\end{tabular}
\end{center}
\caption{
Case of load-dependent driving force $Q=Q(F_\mathrm{ext})$ 
for the illustrative case of Eq.~(\ref{eq:Qlambda}).
}
\label{tab:lambda}
\end{table}

Finally, our calculation can be generalized to the case 
in which the external force is not parallel to the swimming direction of the force-free swimmer. 
For example, for an external force parallel to the $y$ axis, $\F_\mathrm{ext}=F_\text{ext}\e_y$, 
a change in phase speed similar to Eq.~(\ref{eq_force_velocity}) will arise 
if motion in the $x$ and $y$ directions is coupled
with a non-zero hydrodynamic friction coefficient $\Gamma_{xy}$.

\paragraph{Example: three-sphere swimmer.}

We exemplify the above arguments, using Najafi's three-sphere swimmer under external load \cite{Najafi:2004}.
The swimmer consists of three collinear spheres of radius $a$ with respective positions $\mathbf{r}_i$, $i=1,2,3$, 
see Fig.~1A.
While the original model considered a prescribed swimming stroke with constant timing \cite{Najafi:2004}, 
later variants of the three-sphere swimmer considered active driving forces, 
for which the angular frequency depends on the load \cite{Gunther:2008,Felderhof2014_3sphere,Pande:2015}.
Here, we follow the later approach, and impose a non-reciprocal driving protocol with
$d_1=d+A\cos\varphi$, and 
$d_2=d+A\sin\varphi$, 
parametrized by a phase $\varphi$,
where
$\mathbf{r}_2=\mathbf{r}_1+d_1 \mathbf{e}_x$, and
$\mathbf{r}_3=\mathbf{r}_2+d_2 \mathbf{e}_x$. 
The swimmer is immersed in a viscous fluid of dynamic viscosity $\eta$. 
Hydrodynamic interactions between the three spheres are modeled using the Oseen-tensor description \cite{Happel:hydro},
which provides explicit expressions for the hydrodynamic friction coefficients $\Gamma_{ij}$ 
in Eqs.~(\ref{eq_motion}) and (\ref{eq_motion2}) by standard methods \cite{Najafi:2004,Polotzek:2013}.

In the absence of an external force, 
the center $\mathbf{r}=(\mathbf{r}_1+\mathbf{r}_2+\mathbf{r}_3)/3$ of the swimmer moves 
a net displacement $\Delta x$ during one cycle.
The net displacement scales with the square of the amplitude $A$ of the swimming stroke, 
is independent of internal friction,
and reads (to leading order in $a/d$ and $A/d$)
\begin{equation}
\Delta x = \frac{7\pi}{24} \frac{a A^2}{d^2}.
\end{equation}

In the presence of an external force, 
the frequency $\omega$ of the swimming stroke changes to leading order as
\begin{equation}
\omega 
= \omega_0 \left( 1 + \alpha \frac{W}{E} \right), \quad
\alpha=\frac{33+7\ol{\kappa}}{28+7\ol{\kappa}}
, 
\end{equation}
where $\omega_0 = Q / [2\pi \eta a A^2 (4+\ol{\kappa})]$,
$W=\Delta x F_\text{ext}$, and $E=2\pi Q$.
Here, $\ol{\kappa}=\kappa/(\pi\eta a A^2)$ is a dimensionless parameter that characterizes the relative strength of internal friction. 
For the chosen driving protocol, $\dot{\varphi}$ depends slightly on phase $\varphi$, 
and thus the conditions in deriving Eq.~(\ref{eq:omega}) are not strictly fulfilled.
Interestingly, the scaling behavior predicted by Eq.~(\ref{eq:omega}) still applies,
although the numerical prefactor is off by about $20\%$ for $\ol{\kappa}=0$.
We note the swimming efficiency of the swimmer
\begin{equation}
\varepsilon_\mathrm{swim} = \frac{49}{128}\,\frac{a^2 A^2}{(4+\ol{\kappa})\,d^4},
\end{equation}
which yields
$\Phi=\alpha\, \varepsilon_\mathrm{swim}\, \mu F_\text{ext}$.

Fig.~1B shows numerical results for the frequency of the three-sphere swimmer 
as a function of an external force $F_\text{ext}$. 
For the numeric calculations, 
we analytically computed $\dot{\varphi}$ 
by solving the force balance equations, Eqs.~(\ref{eq_motion}) and (\ref{eq_motion2}), 
and then numerically evaluated the integrals 
$T=\oint d\varphi\,\dot{\varphi}^{-1}$
and
$\mu=\oint d\varphi\, \Gamma_{xx}^{-1}\dot{\varphi}^{-1}$.
For small forces, the response is well described by a linear relationship. 
For large forces, nonlinear deviations occur, 
including stalling of the swimming stroke with $\omega(F_\mathrm{stall}^\pm)=0$ 
at critical external forces $F_\mathrm{stall}^\pm$ via a saddle-node bifurcation. 
{\color{black}
For $F_\mathrm{ext}=F_\mathrm{stall}^-<0$ (``head wind''), 
stalling occurs at the peak of the effective stroke with $\phi^-=5\pi/4$,
while for $F_\mathrm{ext}=F_\mathrm{stall}^+>0$ (``tail wind''), 
stalling occurs at the peak of the recovery stroke with $\phi^+=\pi/4$ (independent of $\kappa$).
} 
Note that for these large forces, $|F_\text{ext}|\gg F_a=v_a/\mu$,  
the active contribution to the total swimming speed becomes negligible 
compared to the passive contribution due to the drag by the external force.
The time-averaged hydrodynamic mobility $\mu$ changes only little 
as a function of the external force $F_\text{ext}$, see Fig.~1C.


\begin{figure}[thb]
\begin{center}
\includegraphics[width=\linewidth]{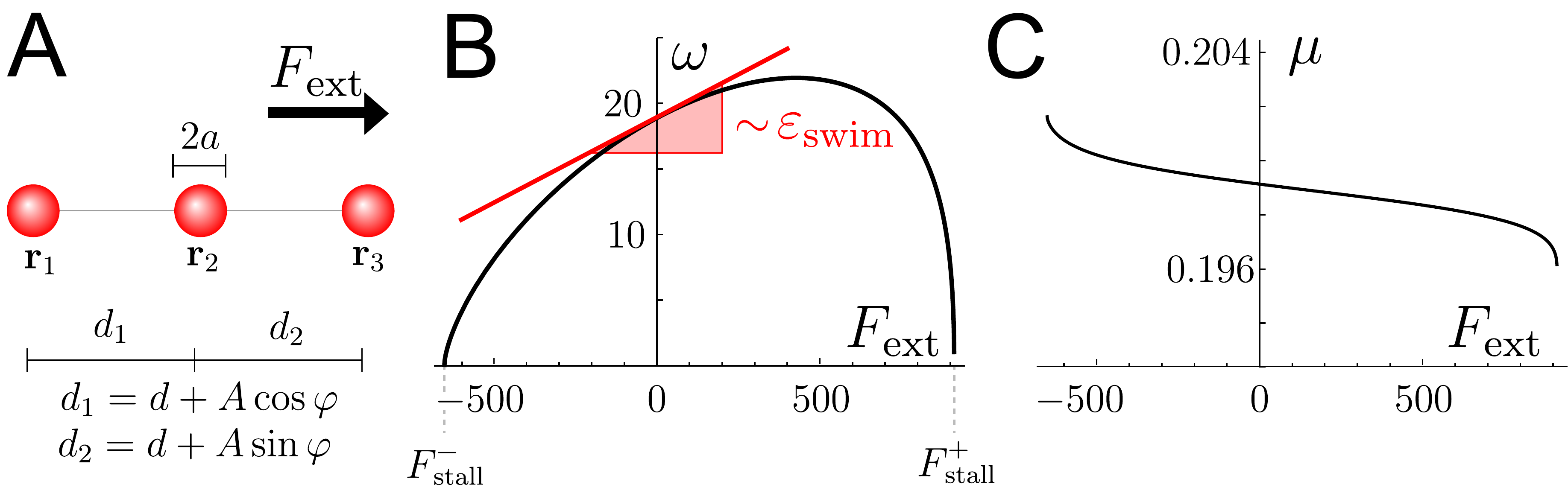}
\end{center}
\caption{
\textit{Three-sphere swimmer under load.}
(A) Schematic representation of the three-sphere swimmer \cite{Najafi:2004}.
(B) Cycle frequency $\omega$ of the swimming stroke as a function of external force $F_\text{ext}$
(black: numerical results, red: analytic theory).
For small external forces, the change in frequency is proportional to the swimming efficiency, $\varepsilon_\mathrm{swim}$.
(C) The time-averaged hydrodynamic mobility $\mu$ as a function of external force $F_\mathrm{ext}$. 
The mobility changes due to a change in the timing of the swimming stroke 
in the presence of an external force, yet this effect is small. 
Frequency in units of $Q/(\eta d^3)$, 
force in units of $Q/d$, 
mobility in units of $1/(\eta d)$.
Parameters: $A/d=0.2$, $a/d=0.1$, $\kappa=0$.
}
\end{figure}

\paragraph{Flagellated swimmers.}

The load response of beating flagella was previously studied 
in the flagellated green algae \textit{Chlamydomonas} \cite{Klindt:2016}, 
which represents a model organism for the study of flagellar self-propulsion.
The internal architecture of cilia and flagella is highly conserved in eukaryotic cells, 
e.g.\ green algae, sperm cells, or ciliated epithelial cells, 
suggesting that similar load responses apply in other cells.  

In the experiments, 
\textit{Chlamydomonas} cells were restrained from moving and exposed to a uniform external flow
$\mathbf{u}=-u\mathbf{e}_y$
that was opposite to the normal swimming direction $\mathbf{e}_y$ of the cell
for unconstrained motion.
It was observed that the flagellar beat accelerated during its effective stroke,
during which the hydrodynamic centers of the two flagella of the cell move in $-\mathbf{e}_y$ direction, 
i.e.\ when the flagella experience ``tail wind'' from the external flow.
Correspondingly, the recovery stroke of the flagellar beat cycle, 
during which the flagella moved against the external flow, slowed down.
For moderate flow speeds ($u<2\,\mathrm{mm/s}$), 
the observed change in the speed of the beat was proportional to the applied flow.
For high flow speeds ($u> 5\,\mathrm{mm/s}$, about $100$ times the unperturbed swimming speed of the cells), 
the flagellar beat stalled in a reversible manner.
The experimental system thus displays signatures 
of the generic load response discussed for the minimal model above:
(i) The instantaneous phase speed $\dot{\varphi}$ of the swimming stroke decreases and increases under load, 
respectively, depending on the direction of instantaneous movement.
(ii) Concomitantly, there is a net change of the cycle frequency $\omega=\langle\dot{\varphi}\rangle$ as a function of load.
(iii) For large external loads, the swimming stroke stalls, corresponding to $\dot{\varphi}=0$. 
Note that in these experiments, also the shape of the flagellar beat changed slightly, 
an effect not considered in our minimal model.

In \cite{Klindt:2016}, the chemo-mechanical efficiency was estimated as
$\varepsilon_\mathrm{chem}\approx 0.2$. 
This value is consistent with direct measurements of the hydrolysis rate of ATP in beating cilia and flagella, 
which correspond to values in the range $\varepsilon_\mathrm{chem}=0.1-0.4$
\cite{brokaw1967adenosine,katsu2009substantial,Cardullo:1991,Chen:2015}.

From an existing hydrodynamic simulation of a swimming \textit{Chlamydomonas} cell \cite{Geyer:2013},
based on flagellar beat pattern obtained from experimental data,
we can compute the time-averaged mechanical power output of a swimming \textit{Chlamydomonas} cell, 
$\langle \mathcal{R}^{(h)} \rangle \approx 25.8\,\mathrm{fW}$, 
as well as the power required to drag the cell through the fluid at its net swimming speed,
$\mathcal{R}_\mathrm{drag}=v_a F_a\approx 0.2\,\mathrm{fW}$.
Here,
$v_a\approx 50.9\,\micron/\mathrm{s}$, 
$\mu\approx 12.2\,\micron/(\mathrm{pN}\mathrm{s})$,
$F_a=v_a/\mu\approx 4.2\,\mathrm{pN}$
for $T=30\,\mathrm{ms}$.
(Similar results were obtained previously with a different numerical method \cite{Klindt:2015}.
Note that the dissipation rate reported there equals $50\%$ 
of the total hydrodynamic dissipation rate of a swimming cell,
corresponding to the mechanical power output of a single flagellum
of the flagellar pair of \textit{Chlamydomonas}.
)
We thus obtain an estimate for the swimming efficiency of a swimming \textit{Chlamydomonas} cell,
$\varepsilon_\mathrm{swim} =   
\varepsilon_\mathrm{hydro}\cdot\varepsilon_\mathrm{chem}
\approx 2\cdot 10^{-3}$, 
with $\varepsilon_\mathrm{hydro} \approx 0.2\,\mathrm{fW}/25.8\,\mathrm{fW} \approx 8\cdot 10^{-3}$.
Previous work reported 
$\varepsilon_\mathrm{swim}  \approx 1\cdot 10^{-3}$  with
$\varepsilon_\mathrm{hydro} \approx 1\cdot 10^{-2}$ 
for multi-ciliated \textit{Paramecium} \cite{katsu2009substantial}, 
and 
$\varepsilon_\mathrm{swim} \approx 1\cdot 10^{-3}$, 
$\varepsilon_\mathrm{swim} \approx 8\cdot 10^{-3}$, 
for demembranated sperm flagella swimming in low and high-viscosity swimming medium, respectively \cite{Chen:2015}.

\paragraph{A superposition principle for multi-component swimmers.}

We now discuss implications for multi-component swimmers consisting of several active components.
This case has been previously considered in \cite{tenHagen2015}, 
yet without taking into account the feedback of external forces on active self-propulsion.
We consider active self-propelled particles that exhibit a generic load response, 
which generalize the shape-changing microswimmers discussed above.

The linearity of the Stokes equation implies a linear relation between
an external applied force $\F$ and torque $\T$ acting on a passive rigid particle immersed in a viscous fluid, 
and the resultant translational velocity $\v$ and rotational velocity $\bm{\Omega}$ of the particle \cite{Happel:hydro}
\begin{equation}
\label{eq:mobility}
\q=
\left(
\begin{matrix}
\v \\ \bm{\Omega}
\end{matrix}
\right)
= 
\left( 
\begin{matrix}
\bm{\mu}_{tt} & \bm{\mu}_{tr} \\
\bm{\mu}_{rt} & \bm{\mu}_{rr} 
\end{matrix}
\right)
\cdot
\left(
\begin{matrix}
\F \\ \T
\end{matrix}
\right)
=\bm{\mu}\cdot\P.
\end{equation}
Here, 
$\bm{\mu}$ denotes the $6\times 6$ hydrodynamic mobility of the particle \cite{Happel:hydro}.
With short-hand 
$\q=(\v;\bm{\Omega})$ and $\P=(\F;\T)$, we can rewrite Eq.~(\ref{eq:mobility}) concisely as $\dot{\q}=\bm{\mu}\cdot\P$.

We now consider a self-propelled particle under the influence of an external force $\P_\text{ext}$. 
The resultant velocity $\q(\P_\text{ext})$ can be written as 
\begin{equation}
\label{eq_force_balance}
\q(\P_\text{ext})=\q_a+\bm{\mu}\cdot\P_\text{ext}+\bm{\Phi}(\P_\text{ext}),
\end{equation} 
which generalizes Eq.~(\ref{eq:phi}) to the case of three-dimensional motion.
In the following, we use a linear expansion 
of the load-response term $\bm{\Phi}$ and neglect higher-order-terms
\begin{equation}
\label{eq_dq}
\bm{\Phi}(\P_\text{ext}) \approx \bm{\chi}\cdot\P_\text{ext}.
\end{equation}
Here, $\bm{\chi}$ denotes a matrix-valued susceptibility.
We introduce the \textit{active mobility} $\bm{\nu}=\bm{\mu}+\bm{\chi}$
and \textit{active friction tensor} $\bm{\Gamma}=\bm{\nu}^{-1}$.

We consider a multi-com\-po\-nent swimmer, consisting of $N$ self-propelled particles 
with respective active velocities $\q_i$ that are connected by a rigid and frictionless scaffold.
The positions $\r_i=\r_0 + r_{ij}\e_j$ of the individual components can be expressed with respect to 
the center $\r_0$ and a material frame with unit vectors $\e_i$, $i=1,2,3$ of the swimmer.
For convenience, the active velocities shall refer to the center of the multi-component swimmer
(using the transformation $\v'_i=\v_i+(\r_i-\r_0)\times\bm{\Omega}_i$,
{\color{black}
where $\v_i$ and $\bm{\Omega}_i$ denote the translational and rotational velocity of the $i$-th component}).

We now ask for the resultant velocity $\q_0$ of the swimmer.
We consider the limit of large separation distances between the individual particles, 
for which hydrodynamic interactions (and possible interference of local fields associated with active propulsion) can be neglected.

Each individual component is subject to a force $\P_i$ exerted by all the other components on it.
Force balance for the whole swimmer implies 
\begin{equation}
\label{eq:Pi}
\sum_i \P_i = 0.
\end{equation}
As each component is moving with velocity $\q_0$, we have by Eq.~(\ref{eq_force_balance})
\begin{equation}
\q_0 = \q_i + \bm{\mu}_i \cdot \P_i + \bm{\chi}_i \cdot \P_i, \quad i=1,\ldots,N .
\end{equation}
Rearranging and summing over all components yields $\sum_i \bm{\Gamma}_i\cdot(\q_0-\q_i)=0$
with $\bm{\Gamma}_i=(\bm{\mu}_i+\bm{\chi}_i)^{-1}$, hence
\begin{equation}
\label{eq_q0}
\bm{\Gamma}_0\cdot\q_0 = \sum_i \bm{\Gamma}_i \cdot \q_i,
\end{equation}
from which $\q_0$ can be computed. 
Here,
\begin{equation}
\label{eq_superposition}
\bm{\Gamma}_0 = \sum_i \bm{\Gamma}_i 
\end{equation}
denotes the active friction tensor of the composite swimmer. 

{\color{black}
Unlike Eq.~(\ref{eq:Pi}), Eq.~(\ref{eq_q0}) does not represent a force balance, 
although the expressions $\K_i=\bm{\Gamma}_i\cdot\q_i$, $i=1,\ldots,N$ and $\K_0=\bm{\Gamma}_0\cdot\q_0$ 
each have units of a (generalized) force. 
Instead, Eq.~(\ref{eq_q0}) restates the superposition principle of low-Reynolds number hydrodynamics \cite{Happel:hydro}, 
which generalizes to multi-component swimmers in the case of a linear force-velocity relation, Eq.~(\ref{eq_dq}).  
The forces $-\K_i$ have a physical interpretation in terms of the constraining force 
required to constrain the $i$-th active component from moving, 
provided the force-velocity relations of its components remain linear also for large forces.
}

\paragraph{Discussion.}

External forces perturb the self-propulsion mechanism of active microswimmers, 
resulting either in an increase or decrease of their active swimming speed.
Here, we discuss the load response of a minimal shape-changing microswimmer.
We show that a change in the frequency of the swimming stroke under load 
scales with the swimming efficiency of the microswimmer
if the energy expenditure per cycle is independent of load.
The swimming efficiency $\varepsilon_\mathrm{swim}$ can be written as a product of 
(i) the Lighthill efficiency $\varepsilon_\mathrm{hydro}$, 
which characterizes the efficacy of active shape changes for self-propulsion 
in terms of hydrodynamic dissipation rates \cite{Lighthill1952}, and 
(ii) a chemo-mechanical swimming efficiency $\varepsilon_\mathrm{chem}$, 
which characterizes the efficiency of the active propulsion mechanism
in executing these shape changes inside a viscous fluid.
The specific example of Najafi's three-sphere swimmer 
has a Lighthill efficiency much smaller than one \cite{Najafi:2004}. 

Our theory implies that the load response of low-efficiency swimmers 
like the three-sphere swimmer is small,
whereas the load response of high-efficiency swimmers 
like the ideal push-me-pull-you 
\cite{Avron:2005} should be large.
(The push-me-pull-you achieves a swimming efficiency above one, 
by pumping an inviscid fluid from one bladder to another 
in a cyclic fashion through a connecting pipe that simultaneously changes its length,
resulting in concomitant cyclic volume changes of the two bladders.)

{\color{black}
The load response of shape-changing microswimmers 
implies that swimming stroke and velocity depend on fluid viscosity, 
as observed e.g.\ for flagellated microswimmers
\cite{Brokaw:1966,Woolley2001,Friedrich:2010}.
It has been proposed that the net swimming speed can even increase for moderate increases in fluid viscosity, 
attaining a maximum at a specific value of fluid viscosity relative to an elastic stiffness of the swimmer
\cite{Krishnamurthy2017,Pande2017}.
A related result was found for a stochastic version of the three-sphere swimmer with
load-dependent rates of conformational changes as a function of an external applied force 
\cite{Golestanian2008conformational}.
Interestingly, for a prescribed swimming stroke without load response, 
the swimming speed would be independent of fluid viscosity \cite{Gray:1955b}.
Artificial microswimmers allow to test these prepositions experimentally \cite{Krishnamurthy2017,Grosjean2016}.
}

{\color{black}
Similar results should hold also for shape-changing microswimmers in non-Newtonian fluids. 
The forces exerted by the fluid on the swimmer causing a load response can be either
pure hydrodynamic friction forces as for the case of a Newtonian fluid considered here, 
or, more generally, a combination of elastic and viscous forces in the case of a non-Newtonian fluid
\cite{Lauga2009,Qiu2014}. 
}

In addition to shape-changing microswimmers,
we expect that similar load responses hold also for diffusio-phoretic swimmers.
There, local concentration gradients of a solute, 
generated by chemical reactions that are asymmetrically distributed on the surface of the swimmer,
drive active motion of the swimmer due to different surface mobilities of solute and solvent 
\cite{Anderson1989,Golestanian2005,Julicher2009}.
External forces that passively drag such a phoretic swimmer in addition to its active propulsion
will distort the concentration fields of the solute in the vicinity of the microswimmer, 
and thus change its active propulsion velocity. 
Michelin {et al.} calculated the effect of finite Peclet number 
on phoretic self-propulsion \cite{Michelin2014}.
Similar methods should allow to address the active swimming velocity in the presence of an external force.

In conclusion, 
we introduced a minimal model for fluid-structure interactions
between active, shape-changing structures and a viscous fluid, 
which has implications for our understanding of active propulsion and multi-component swimmers.

\paragraph{Acknowledgment.}
The author is supported by the DFG through the 
Excellence Initiative by the German Federal and State Governments (cluster of excellence cfaed), 
as well as the ``Microswimmers'' priority program (FR3429/1-1, FR3429/1-2).

\bibliography{../../bibliography/library,./superposition}

\end{document}